\documentclass[aps,prb,twocolumn,superscriptaddress,showpacs]{revtex4}
\usepackage{graphicx}

\newcommand{\tabrule}{\rule[-0.2em]{0em}{1.2em}}
\usepackage{amsmath}
\usepackage{booktabs}
\usepackage{color}

\begin{document}

\title{Magnon-driven longitudinal spin Seebeck effect in  $F|N$ and $N|F|N$ structures: role of asymmetric  in-plane magnetic anisotropy }

\author{L. Chotorlishvili\footnote{levan.chotorlishvili@gmail.com}}
\affiliation{Institut f\"ur Physik, Martin-Luther-Universit\"at Halle-Wittenberg,
06099 Halle, Germany}
\author{Z. Toklikishvili}
\affiliation{Department of Physics, Tbilisi State University, Chavchavadze av. 3, 0128, Tbilisi, Georgia}
\author{S. R. Etesami}
\affiliation{Institut f\"ur Physik, Martin-Luther-Universit\"at Halle-Wittenberg,
06099 Halle, Germany}
\affiliation{Max-Planck-Institut f\"ur Mikrostrukturphysik, Weinberg 2, 06120 Halle, Germany}
\author{V. K. Dugaev}
\affiliation{Institut f\"ur Physik, Martin-Luther-Universit\"at Halle-Wittenberg,
06099 Halle, Germany}
\affiliation{Department of Physics, Rzesz\'{o}w University of Technology,
al. Powstanc\'ow Warszawy 6, 35-959 Rzesz\'ow, Poland}
\affiliation{Departamento de F\'isica and CFIF, Instituto Superior T\'ecnico,
Universidade de Lisboa, Av. Rovisco Pais, 1049-001 Lisbon, Portugal}
\author{J. Barna\'s}
\affiliation{Faculty of Physics, Adam Mickiewicz University, ul. Umultowska 85, 61-614 Pozna\'n, Poland}
\affiliation{Institute of Molecular Physics, Polish Academy of Sciences, ul. Smoluchowskiego 17, 60-179 Pozna\'n, Poland}
\author{J. Berakdar}
\affiliation{Institut f\"ur Physik, Martin-Luther-Universit\"at Halle-Wittenberg,
 06099 Halle, Germany}

\date{\today}

\begin{abstract}
The influence of an asymmetric in-plane magnetic anisotropy $K_{x}\neq K_{y}$ on the thermally activated spin current is studied theoretically for two different systems; (i) the $F|N$ system consisting of a ferromagnetic insulator ($F$) in a direct contact with a nonmagnetic metal ($N$),
and (ii) the sandwich structure $N|F|N$ consisting of a ferromagnetic insulating part sandwiched between two nonmagnetic metals.
It is shown that when the difference between the temperatures of the two nonmagnetic metals in a $N|F|N$ structure is not large, the spin pumping currents from the magnetic part to the nonmagnetic ones are equal in amplitude and have  opposite directions, so only the spin torque current contributes to the total spin current. The spin current flows then from the nonmagnetic metal with the higher temperature to the nonmagnetic metal having a lower temperature. Its amplitude varies linearly with the difference in temperatures. In addition, we have  found that if the magnetic anisotropy is in the layer plane, then the spin  current increases with the magnon temperature, while in the case of an out-of-plane magnetic anisotropy the spin  current decreases when the  magnon temperature enhances.
Enlarging the difference between the temperatures of the nonmagnetic metals,
 the linear response becomes important, as confirmed by  analytical expressions inferred from the Fokker-Planck approach and by the results obtained upon a full numerical integration of the stochastic Landau-Lifshitz-Gilbert equation.
\end{abstract}

\maketitle

\section{Introduction}

One of the key observations that gave impetus to the field of  spin caloritronics was the discovery of the spin
Seebeck effect (SSE) \cite{Uchida2008} which amounts to the emergence of  a spin current
upon an externally applied thermal
gradient\cite{Hatami2009,Bosu2011,Jaworski2010,Uchida2010,Xiao2010,Uchida-Nonaka2010,Sears2012,Torrejon2012,
Li2012,Jia2011,Adachi2013,Tserkovnyak2002,Chotorlishvili2013,Etesami,Hoffman2013,Agrawal2013,Kikkawa2013}.
Technologically, various  SSE-based  nanoelectronics devices are envisaged.
For example, portable thermal diodes have been proposed to control and rectify the heat and spin
currents.

The SSE was observed in materials of substantially different transport properties such as  metallic ferromagnet Co$_2$MnSi, semiconducting
ferromagnet GaMnAs, and magnetic insulators LaY$_{2}$Fe$_{5}$O$_{12}$ and  (Mn, Ze)Fe$_{2}$O$_{4}$. Thus, the underlying mechanism
may well depend on the specific case under study as well as on the experimental setup.
In metallic ferromagnetic systems, the spin is transferred {\it via}  charge carriers activated by
the thermal bias, while in magnetic insulators the SSE
is mediated by magnons flowing towards the cold edge of the sample.
A theory of the magnon driven SSE was developed in Ref.~\onlinecite{Xiao2010}
and  implemented beyond the linear response regime. \cite{Chotorlishvili2013,Etesami}
Thereby, the concept of magnon temperature is of a  key importance for understanding the physical
origin of the spin current in magnetic insulators: An external heat bias applied to the system
thermalizes the phonon subsystem much faster than the magnons relax. Therefore, the magnon
temperature $T^m_F$ is influenced by the already established phonon temperature profile $T_F$.
The thermally activated spin current is related to the
temperature difference between the phonon and magnon subsystems.

Recent studies based on the macrospin
approach (valid for samples of small dimensions on the range of the exchange length) concern the linear  \cite{Xiao2010},
and  nonlinear response regimes \cite{Chotorlishvili2013}. In both regimes, the
obtained analytical expressions for the spin current are proportional to the difference between
the phonon and magnon temperatures. As shown in Ref.~\onlinecite{Chotorlishvili2013}, the result for the spin current obtained
in the linear response theory is a particular case of the result obtained
in the Fokker-Planck approach, and corresponds to the low magnon temperature regime.
The nonlinear effects substantially change the role of the magnetic anisotropy
in the formation of the spin current. In the linearized approach, the role of the magnetic anisotropy is
similar to that of an external magnetic field, and can be described by a certain effective field.
This is not the case when magnetic fluctuations are large which is more likely  for higher temperatures.

A quantitative criterion for the threshold
magnon temperature, above which the anisotropy plays an important role, is defined by the following inequality\cite{Chotorlishvili2013}, $T^m_F>M_sVH_{\rm eff}/k_B$.
Here $M_s$ is the saturation magnetization, $V$ is the total volume of the ferromagnet, and $H_{\rm eff}$
is the effective magnetic field.  In the present work we
focus on  phenomena inherent to the nonlinear regime. In particular, we study
the influence of the in-plane magnetic anisotropy  on the SSE. We show that the
effect of the in-plane anisotropy on the thermally activated spin current is different from the effect
of the uniaxial out-of-plane anisotropy studied in Ref.~\onlinecite{Chotorlishvili2013}.

We consider two different system relevant for the longitudinal SSE: (i) an $F|N$ structure consisting of a ferromagnetic insulating part ($F$) attached to a nonmagnetic metallic part ($N$), and (ii) a ferromagnetic insulating part sandwiched between two nonmagnetic metallic ones ($N|F|N$). In the following the nonmagnetic metallic part will be referred to as metallic part or simply as metal, while the ferromagnetic insulating part will be referred to as  ferromagnetic part or simply as ferromagnet. We show that in the both cases thermally activated spin current is parallel to the temperature gradient. In the case of $N|F|N$ structure we show that if the difference between the temperatures of the two metals is not large, and the temperature dependance of the spin conductance can be ignored, then the spin pumping currents from the ferromagnet to the adjacent metals are equal in amplitude and are oriented oppositely, so they do not contribute to the total spin current. The only contribution to the total spin current is then due to the spin torque current. We show that the spin torque current flowing through the ferromagnetic part is a linear function of the difference between the temperatures of the two metals. Though, the considerations apply directly to the case when the ferromagnetic part is insulating, the results are also applicable when the ferromagnetic part is metallic.

In Section 2 we present the model $F/N$ structure. The Fokker-Planck approach is briefly described  in Section 3,
where analytical results for the spin current in the $F|N$  system are  presented. Numerical results for the total spin current in the $F|N$ system,
obtained by a direct numerical integration of the stochastic Landau-Lifshitz-Gilbert (LLG) equation, are presented and discussed in Section 4.
We show that in the presence of an in-plane magnetic anisotropy and a weak magnetic field, the ($F|N$) system supports a spin pumping current only if the axial symmetry in the system is broken by an in-plane magnetic anisotropy.
In Section 5 we discuss analytical results on the spin current flowing through the system $N|F|N$, while the corresponding
results obtained by numerical integration are presented in Section 6.
We show that such a system supports spin pumping current only for a sufficiently large thermal gradient.
Summary and final conclusions are in Section 7.

\section{Model of the $F|N$ structure}

We consider first the thermally activated spin current  through the
interface  in the $F|N$ system. Our main focus is on the influence of the in-plane
magnetic anisotropy, $-M_{s}\big(K_xm_x^2+K_ym_y^2\big)/2$, on the
spin current. Here, $\bf m$ is a unit vector along the magnetic moment, $\vec{m}=\vec{M}/M_{s}$.
We assume that the thermal equilibrium between the electron and the phonon subsystems in the metallic part
as well as in the ferromagnetic part is restored internally  much faster than the equilibrium between the two parts.
In terms of the local temperature, which is based on the hierarchy of relaxation
times, this means that the temperatures of the phonons, $T^p_{N(F)}$,  and the
electrons $T^e_{N(F)}$, baths are equal in both the metallic and
ferromagnetic parts, $T^p_N=T^e_N=T_N$, $T^p_F=T^e_F=T_F$. However, there is
a difference in temperatures of the two components, $T_F\neq T_N$,  which can be controlled externally.
This difference drives the SSE. The interaction between the nonmagnetic  and ferromagnetic subsystems
is mediated {\it via} the magnon bath, described by the magnon temperature $T_F^m$. Due to the slower magnon relaxation, $T_F^m$ may be different from $T_F$.

\begin{center}
   \begin{figure}[h]
    \centering
    \includegraphics[width=0.96\columnwidth]{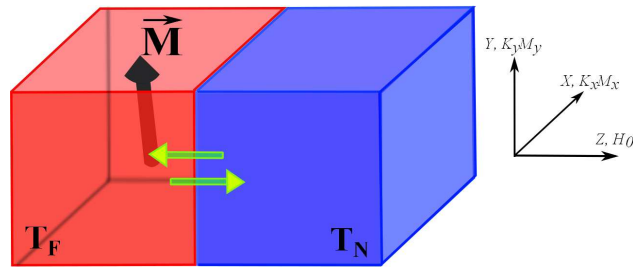}
        \caption{\label{skim} A schematic illustration of the considered $F|N$ system. The system consists of a ferromagnetic insulator (left) and a nonmagnetic metal (right)
         in  direct contact. The temperatures of the ferromagnetic ($T_F$) and metallic ($T_N$) parts are in general different.}
    \end{figure}
\end{center}
The magnetization dynamics of the ferromagnetic part, to be considered in the following section, is described by the
LLG equation in the macrospin approximation. Such an approach excludes nonuniform magnetization, and therefore is applicable
when the magnetic component of the system has a small volume, i.e., its lateral and vertical dimensions are small, usually in the nanometer range. Using the Fokker-Plank-equation technique, we evaluate the
mean value of the spin current flowing through the interface from the ferromagnetic part to the metallic one.
We also study there
the dependence of the total spin current on the in-plane magnetic anisotropy.

\section{Fokker-Planck approach: analytical solution for the spin current in the $F|N$ system}

The total spin current flowing through the interface in the $F|N$ system
consists of two contributions -- the spin pumping current flowing from the ferromagnet to the
normal metal, and the thermally activated spin current flowing in the opposite direction (in the following also referred to as the spin torque current). Powered by the phonon bath, thermal noise leads to formation of the fluctuating spin torque current
in the normal metal, $\vec{I}_{fl}(t)$. Effect of this fluctuating spin torque current on the ferromagnetic insulator can be described by a random magnetic field $\vec{h}'(t)$ acting on the magnetization,\cite{Xiao2010} $\vec{I}_{fl}(t)=-\frac{M_{s}V}{\gamma}\gamma\vec{m}(t)\times\vec{h}'(t)§$. Here $M_s$ is the saturation magnetization, $V$ is the total volume of the ferromagnet, and $\gamma$ is the gyro-magnetic factor. On the other hand, the thermally activated magnetization dynamics in the ferromagnet gives rise to a spin pumping current
emitted from the ferromagnet into the normal metal, $\vec{I}_{s}=\frac{M_{s}V}{\gamma}\alpha'\vec{m}\times\dot{\vec{m}}$. Thus, the total
average {\it dc} spin current across  the interface can be written in the form \cite{Xiao2010,Chotorlishvili2013}
\begin{equation}
\label{eq_1}
    \langle\vec{I}_s\rangle=\frac{M_sV}{\gamma}\left[\alpha'\langle\vec{m}
		\times\dot{\vec{m}}\rangle-\gamma\langle\vec{m}\times\vec{h}'\rangle\right] ,
\end{equation}
where , $\alpha'=(\gamma\hbar/4\pi M_sV)g_r$ is the magnetization
damping constant (related to the spin pumping), and $g_r$ is the
real part of the dimensionless spin mixing conductance.
Furthermore, $\vec{h}'(t)$ is the random magnetic field. If the random thermal force has correlation time much shorter than the response time of the system, one can assume that the noise is white. A quantitative criterion for using a white noise is~\cite{Xiao2010} $k_{B}T\gg \hbar \omega_{0}$, where  $\omega_{0}$ is the ferromagnetic resonance frequency. Taking into account the fact that the response time
of the system is defined by $\omega_{0}\approx$ 1 GHz, for the low temperature limit one obtains $T>>0.01$K. Evidently, this temperature regime allows using the white noise in our problem \cite{Brawn,Trimper}:
\begin{equation}\label{eq.3}
\langle\gamma h'_{i}(t)~\gamma h'_{j}(t')\rangle
=\sigma'^{2}\delta_{ij}\delta(t-t'),
\end{equation}
for $i,j=x,y,z$,
where $\sigma'^{2}=2\alpha'\gamma k_BT_N/M_sV$.
Thus, to find the spin current we need
to know the time evolution of the magnetization.
The magnetization dynamics is described by the stochastic LLG equation
for the dimensionless unit vector $\vec{m}$
\begin{equation}
\label{eq_2}
    \dot{\vec{m}}=-\gamma\vec{m}\times\left(\vec{H}_{\rm eff}+\vec{h}\right)+\alpha\vec{m}
		\times\dot{\vec{m}}.
\end{equation}
The effective magnetic field $\vec{H}_{\rm eff}$ consists of the external field applied
along the $z$ axis and of the in-plane magnetic
anisotropy field $\vec{h}_A=-\partial V_{\rm an}/\partial\vec{M}=K_{x}m_{x}\vec{i}_{x}
+K_{y}m_{y}\vec{i}_{y}$, where $\vec{i}_{x(y)}$ are the unit vectors along the axis $x$($y$).
The magnetic anisotropy energy density,
$V_{\rm an}=-\frac{1}{2}M_{s}(K_xm_x^2+K_ym_y^2$), is described by the anisotropy
constants $K_{x}$ and $K_{y}$.
In the above equation, $\alpha$ is the total magnetic damping constant,
$\alpha T_{F}^{m}=\alpha_{0}T_{F}+\alpha'T_{N}$. This  damping constant includes
the contributions from the standard bulk damping constant $\alpha_{0}$ which is associated with the
lattice thermal oscillations (Gilbert damping) and from the damping constant $\alpha'$ associated
with the contact to the normal metal. Introducing the enhanced total damping constant $\alpha$ has a clear physical motivation. Due to the effect of the $F|N$ interface, magnetization dynamics in the F layer is additionally damped, and this enhanced damping is due to a magnonic spin current transferred from the ferromagnetic insulator to the normal metal.\cite{Kapelrud}  Also here we assumed that the random contributions from the uncorrelated noise sources are totally independent and, therefore, the total enhanced damping constant is factorized \cite{Xiao2010}.  Finally, $\vec{h}$  in Eq.(3)  is the total
random field, with the corresponding
correlation function of the form \cite{Xiao2010}
\begin{equation}
\label{eq_3}
    \langle h_i(t) h_j(t^\prime)\rangle=\sigma^2\delta_{ij}\delta(t-t^\prime),
\end{equation}
where $\sigma^2=2\alpha k_BT_F^m/M_sV\gamma$ is the coefficient  proportional to the
magnon temperature.

Following the procedure described in Ref.~\onlinecite{Chotorlishvili2013}, we derive
the Fokker-Planck (FP) equation
for the distribution function $f(\vec{m},T)$:
\begin{equation}
\label{eq_4}
\begin{split}
 &\frac{\partial f}{\partial t}=\frac{1}{1+\alpha^2}\frac{\partial}{\partial \vec{m}}
\Bigg\{ \left(\vec{m}\times\vec{\omega}_{\rm eff}\right)f\\ &+\alpha\vec{m}\times\left(\vec{m}
\times\vec{\omega}_{\rm eff}\right)f-\frac{\sigma^2}{2(1+\alpha^2)}\vec{m}\times\left(\vec{m}
\times\frac{\partial f}{\partial \vec{m}}\right)\Bigg\} ,
  \end{split}
\end{equation}
where $\vec{\omega}_{\rm eff}=\gamma \vec{H}_{\rm eff}=\left(\omega_xm_x,\omega_ym_y,\omega_0\right)$.
The stationary solution of the FP equation reads:
\begin{equation}
\label{eq_5}
    f(\vec{m})=Z^{-1}\exp\left(\beta\int\vec{\omega}_{\rm eff}.d\vec{m}\right) ,
\end{equation}
where $Z=\int\exp\left(\beta\int\vec{\omega}_{\rm eff}\cdot d\vec{m}\right)d^3\vec{m}$ is the normalization
factor, and we introduced the following notation: $\beta=2\alpha\left(1+\alpha^2\right)/\sigma^2$.
Taking into account Eqs.~(\ref{eq_1}) to (\ref{eq_5}), after some cumbersome, but straightforward  calculations
we find the following expression for the  average total spin current:
\begin{equation}
\label{eq_6}
\begin{split}
 \langle I_{sz}\rangle=\alpha'k_B\Big\{&T_F^mA\langle1-m_z^2\rangle-2T_F^mB_x\langle m_zm_x^2\rangle\\
  &-2T_F^mB_y\langle m_zm_y^2\rangle-2T_N\langle m_z\rangle\Big\},
  \end{split}
\end{equation}
where $A=\frac{M_sVH_0}{k_BT_F^m}$, $B_x=\frac{K_xM_sV}{2k_BT_F^m}$, and $B_y=\frac{K_yM_sV}{2k_BT_F^m}$,
while the averages occurring in Eq.~(\ref{eq_6}) are defined in the Appendix [see Eqs.~(\ref{eq_A2}) to (\ref{eq_A6})].
By definition, the spin current is generally a second rank tensor and is characterized by spatial orientation and projection of momentum,  $I_{x,y,z}^{M_{x,y,z}}$. However, due to the particular geometry under consideration, only the longitudinal spin current component, $I_{z}^{M_{z}}=I_{sz}$, is nonzero, while the transversal spin current components vanish,  $I_{sx}=I_{sy}=0$. Thus our findings are in favor of the recent experiment \cite{Srichandan} in which
upper limit for the transverse spin Seebeck effect was observed several orders of magnitude smaller than previously reported.
In spite of the fact that expression of the spin current doesn't depend on the enhanced damping constant $\alpha$ explicitly, due to the
relation $\alpha T_{F}^{m}=\alpha_{0}T_{F}+\alpha'T_{N}$ spin current still depends on $\alpha$ implicitly. One can invert dependence on
the spin mixing conductance damping $\alpha'$ into the dependence on $\alpha$. However precise values of $\alpha'$ is too much related
to the characteristics of particular interface. Therefore for the sake of general interest we quantify spin current in terms of $<I_{sz}>/\alpha'k_{B}$.
The in-plane magnetic anisotropy has different physical consequences, when compared to the case of an out-of-plane anisotropy studied in Ref. \onlinecite{Chotorlishvili2013}. First of all, the expression for the total spin current in the case of an in-plane anisotropy is different from that for an out-of-plane anisotropy. Only in the case of an axial symmetry, $K_{x}=K_{y}$, the expressions for the total spin current become  partly similar. However, as we will see below, the main  effect of the in-plain anisotropy concerns the asymmetric case, $K_{x}\neq K_{y}$.

The expression for the total spin current, Eq.~(\ref{eq_6}),
is quite general. Therefore, we consider now some asymptotic situations. In the symmetric case,
$K_x=K_y=H_{A}$, one finds $B_x=B_y=B=H_{A}M_sV/2k_BT_F^m$.  Equation~(\ref{eq_6}) reduces then to a simpler form, while the mean components
of the magnetization can be calculated analytically:
\begin{subequations}
\label{eq_7}
\begin{align}
 &\langle m_z\rangle= \frac{A}{2B}+\frac{2}{\sqrt{\pi B}}\frac{e^{-\frac{A^2}{4B}-B}\sinh[A]}{G(A,B)},\\
 &\langle1-m_z^2\rangle = 1-\left(\frac{A}{2B}\right)^2-\frac{1}{2B}\nonumber \\
 &-\frac{e^{-\frac{A^2}{4B}-B}(2B\cosh(A)+A\sinh(A))}{G(A,B)},\\
 &\langle m_z\left(m_x^2+m_y^2\right)\rangle
 = -\left(\frac{A}{2B}\right)^3-\frac{3}{4}\frac{A}{B^2}+\frac{A}{2B} \nonumber \\
 &-\frac{e^{-\frac{A^2}{4B}-B}}{2\sqrt{\pi}B^{5/2}G(A,B)} \nonumber \\
&\times \left[2AB\cosh(A)+\left(A^2+4B\right)\sinh(A)\right].
\end{align}
\end{subequations}
Here, we introduced the following notation:
$G(A,B)={\rm erf}\left((A-2B)/2\sqrt{B}\right)-{\rm erf}\left((A+2B)/2\sqrt{B}\right)$,
where  ${\rm erf}(\cdot\cdot\cdot)$ is the error function.
Interestingly, the in-plane magnetic anisotropy in the symmetric case
is equivalent to the out-of-plane anisotropy,
$-H_{A}m_{z}^{2}$. Therefore, after substituting $B\rightarrow -B$ in
Eqs~(\ref{eq_7}), we recover the results derived earlier \cite{Chotorlishvili2013}.

In the case of weak magnetic anisotropy, $B<1$, the mean components of the magnetic moment can
be further simplified (see Eq.~(\ref{eq_A7}) to Eq.~(\ref{eq_A10}) in the Appendix).
In turn, in the absence of magnetic
anisotropy, $B\rightarrow0$, from Eqs.~(\ref{eq_6}) and
(\ref{eq_7}) one finds $\langle m_z\rangle=L(A)$,
$\langle(1-m_z^2)\rangle=2L(A)/A$, and $\langle I_{sz}\rangle=2\alpha'
k_BL(A)(T_F^m-T_N)$, where $L(A)=\coth(A)-1/A$ is the
Langevin function. This result (obtained for
$B\rightarrow0$) recovers the previously obtained expression for the total
spin current \cite{Chotorlishvili2013}.

Now we address the case of a high magnon temperature, $M_sVH_0/k_BT_F^m<1$ and
$H_{A}M_sV/2k_BT_F^m<1$. Then, from Eqs.~(\ref{eq_6}) and (\ref{eq_7})
we obtain $\langle I_{sz}\rangle=2\alpha'
k_B\frac{M_sVH_0}{3k_BT_F^m}\left(T_F^m-T_N\right)\left(1-\frac{2}{3}\frac{H_{A}M_sV}{2k_BT_F^m}\right)$ for the total spin current.
From this formula follows that the positive in-plane anisotropy ($H_{A}>0$) reduces the total spin
current, while the negative in-plane anisotropy ($H_{A}<0$) enhances the total spin current.
Apart from this, the spin
current is proportional to the difference $\left(T_F^m-T_N\right)$
between the magnon temperature and the temperature of the  metal.
The amplitude of the total spin current, however, is reduced because of the
anisotropy term $\frac{2}{3}\frac{H_{A}M_sV}{2k_BT_F^m}$. Another interesting
observation is that the spin current
vanishes, $\langle
I_{sz}\rangle\rightarrow0$, for zero magnetic field, $H_0\rightarrow0$. This result is rather clear since
the dynamics of the magnetization is then strongly exposed to
the thermal fluctuations, and therefore suppresses the spin current generation.
\begin{center}
   \begin{figure}[t!]
    \centering
    \includegraphics[width=0.95\columnwidth]{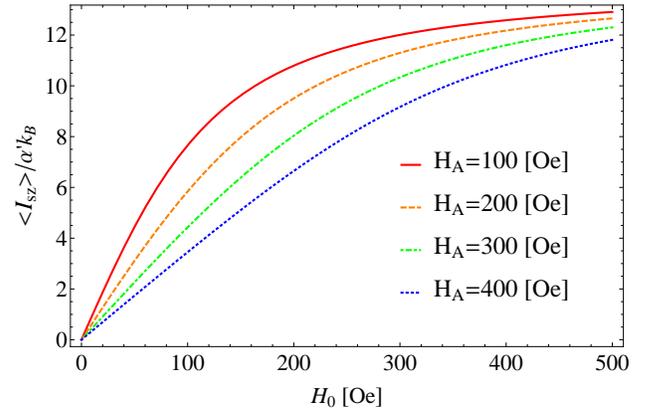}
        \caption{\label{fig1} Dependence of the average total spin current
				$\langle I_{sz}\rangle/\alpha^\prime k_B$ on the external magnetic field $H_0$
				for the following parameters:  $K_x=K_y=H_A>0$ (with $H_A$ as indicated), $T_F^m=300$~K,
				$T_N=297$~K, $M_s=800$~G, and $V=1.6\times10^{-18}$~cm$^3$. The figure shows that  the spin current decreases with the increasing the magnetic anisotropy. Since we measure spin current in the units of $\langle I_{sz}\rangle/\alpha^\prime k_B$ spin current doesn't depend on the
$\alpha^\prime$. Besides analytical result is obtained via the steady sate solution of the FP equation. This implicates that system is already relaxed. This is reason why spin current doesn't depend on the $\alpha$ as well. Values of the inverse damping constant $1/\alpha$ defines relaxation rate and
steady sate distribution function is formed beyond this time scale.}
    \end{figure}
\end{center}

In the case of a strong magnetic field and a weak anisotropy,
$\frac{1}{2}H_{A}M_sV<k_BT_F^m<H_0M_sV$, we enter a different regime,
and the expression for the total spin current in this regime reads
\begin{equation}
\label{eq_8}
 \langle I_{sz}\rangle=2\alpha' k_B\big(T_F^m-T_{N}\big)\bigg(1-\frac{3H_{A}}{H_0}
\frac{k_BT_F^m}{M_sVH_0}\bigg).
\end{equation}
The first term in Eq.~(\ref{eq_8}) is the standard contribution
to the spin current in the linear response, while the second term is
due to the magnetic anisotropy. As before, the positive in-plane
anisotropy  ($H_{A}>0$) suppresses the total spin
current, while the negative in-plane anisotropy ($H_{A}<0$) enhances the spin current.
\begin{center}
   \begin{figure}[t!]
    \centering
    \includegraphics[width=0.95\columnwidth]{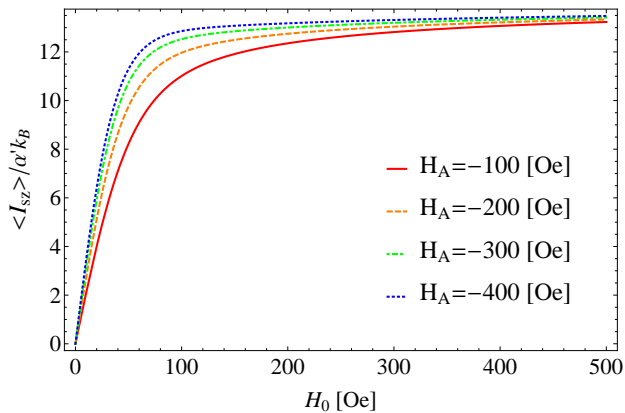}
        \caption{\label{fig2} The average total spin current  as function of the external magnetic field $H_0$
                for the following parameters: $K_x=K_y=H_A<0$ (with $H_A$ as indicated), $T_F^m=300$~K,
				$T_N=297$~K, $M_s=800$~G, and $V=1.6\times10^{-18}$~cm$^3$. Absolute values of the magnetic
				anisotropy field are the same as in Fig.~\ref{fig1}, but here they have the opposite sign.
				Note, the spin current now increases with increasing $|H_A|$.}
    \end{figure}
\end{center}

Having considered  some limiting asymptotics for the thermally activated spin current, let us go back to the general solution, see  Eqs.~(\ref{eq_6}) and
(\ref{eq_7}). Since the general solution is relatively complex for its illustration we plot the
total average spin current as a function of the external magnetic
field $H_0$ and  the magnon temperature $T_F^m$. First, we consider a symmetric situation, $K_x=K_y=H_{A}$, and then
proceed with more important an asymmetric case, $K_x\ne K_y$.

\begin{center}
   \begin{figure}[]
    \centering
    \includegraphics[width=0.95\columnwidth]{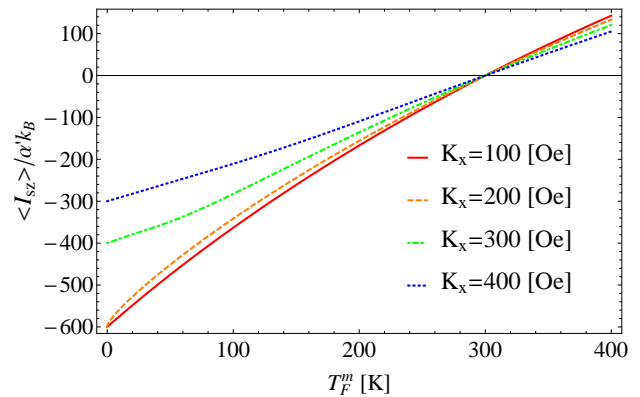}
        \caption{\label{fig5} The average total spin current  as  the magnon
				temperature $T_F^m$  varies. The following parameters are chosen: $K_y=100$~Oe,
				$H_0=200$~Oe, $T_N=300$~K, $M_s=800$~G, $V=1.6\times10^{-18}$~cm$^3$, and $K_x$ as indicated.}
    \end{figure}
\end{center}

Figures \ref{fig1} and \ref{fig2} show the dependence of the total spin current on
the external magnetic field $H_0$ applied along the $z$ axis. As one can note, the magnitude
of spin current depends on the sign of the anisotropy field $H_A$. For positive
magnetic anisotropy, Fig.~\ref{fig1}, the spin current decreases with
increasing the anisotropy field, while for negative anisotropy, the
spin current grows with increasing absolute value of $H_A$, see Fig.~\ref{fig2}.
In turn, the dependence of the average spin current on the magnon temperature is presented in Fig.\ref{fig5} for the indicated values of the
anisotropy constant $K_x$ and the constant $K_y$. All curves cross at the point $T_F^m=T_N$, where the spin current is equal to zero as the system is then in thermal equilibrium. Interestingly, the spin current for $K_x>K_y$ is reduced in comparison to the spin current in the symmetrical case ($K_x=K_y$). This is because the $x$-component of magnetization increases with the increasing constant $K_x$ (positive), and thus the magnetization projection on the $z$-axis becomes reduced.

\begin{center}
   \begin{figure}[t!]
    \centering
    \includegraphics[width=0.95\columnwidth]{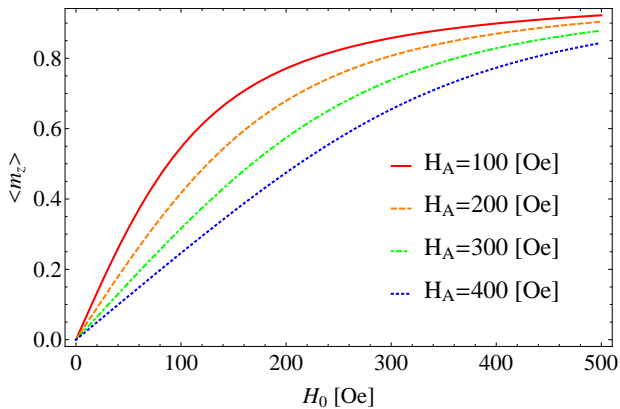}
        \caption{\label{figMH} Dependence of the averaged component of magnetization $\big<m_{z}\big>$ on the external magnetic field $H_0$
				for the following parameters:  $K_x=K_y=H_A>0$ (with $H_A$ as indicated), $T_F^m=300$~K,
				$T_N=297$~K, $M_s=800$~G, and $V=1.6\times10^{-18}$~cm$^3$. The magnetic field $H_{0}=(0,0,H_{0})$ tends to align
               magnetization vector along the $Z$ axis, while the in-plane magnetic anisotropy
               favors the in-plane alignment of the magnetization.}
    \end{figure}
\end{center}

Before proceeding to a more important case of asymmetric in-plane anisotropy ($K_x \neq K_y$), we plot the dependence of the
mean magnetization component $\big<m_{z}\big>$ on the applied magnetic field. As follows from Fig.~\ref{figMH}, the magnetization component $\big<m_{z}\big>$ increases with the applied magnetic field, approaching the  maximum $\big<m_{z}\big>\approx 1$.
However, the limit $\big<m_{z}\big>= 1$ corresponds to the zero magnon temperature,
$A=\frac{M_sVH_0}{k_BT_F^m} \rightarrow \infty,~~T_F^m=0$, and therefore is beyond the
Fokker-Plank approach.

Finally, let us consider the limit of low magnon temperature.  Taking into account Eq.~(\ref{eq_5}), Eq.~(\ref{eq_6}),
and applying the saddle-point method, we find the following formula for the average total spin
current in the limit of a weak magnetic field
$2H_0/\left(K_x+K_y\right)<1$:
\begin{equation}
\label{eq_9} \langle I_{sz}\rangle=-\alpha'
k_B\eta_1\Bigg\{\eta_2\frac{I_1(\eta_2/T_F^m)}{I_0(\eta_2/T_F^m)}+T_N\Bigg\}.
\end{equation}
Here, we introduced the following notation:
$\eta_1=H_0/\left(K_x+K_y\right)$ and
$\eta_2=\frac{M_sV}{4k_B}\left(K_x-K_y\right)\left(1-4\eta_1^2\right)$.
In the opposite case of $2H_0/\left(K_x+K_y\right)>1$,
the spin current is given by the expression $\langle I_{sz}\rangle=-2\alpha' k_BT_N$.

The second term in Eq.~(\ref{eq_9}) corresponds
to the spin torque current flowing from the  metal to the ferromagnet, so it is negative,
$-\alpha'k_B\eta_1T_{N}$. Taking into account the parity of Bessel functions
$I_1(\eta_2/T_F^m)$ and $I_0(\eta_2/T_F^m)$, it is easy to see
that the first term $-\alpha'k_B\eta_1 \eta_2 I_1(\eta_2/T_F^m)/I_0(\eta_2/T_F^m)$ is even with
respect to the
permutation $(K_{x}-K_{y})\rightarrow -(K_{x}-K_{y})$, and is always negative. It disappears
in the symmetric case of $K_{x}=K_{y}$ while in the antisymmetric case
it decreases the spin torque current, see Fig.~\ref{fig5} in the low magnon temperature regime.

\section{Numerical solutions for the $F|N$ system}

The results derived analytically in the previous section from the solution of
the Fokker-Planck equation can be further confirmed by exact
numerical integration of the stochastic LLG equation,
assuming the random field $\vec{h}$ as
a Gaussian white noise defined through the correlation function, see Eq.~(\ref{eq_3}).
The magnon temperature $T_F^m$ is implemented into the simulations {\it via} the strength parameter of
the correlation function, $\sigma^2=2\alpha k_BT_F^m/M_sV\gamma $ (see also section 3).

To solve the stochastic LLG equation we used the Heun method, which converges in the quadratic
mean to the solution interpreted in the sense of Stratonovich \cite{Kloeden}.
From the solutions of  the stochastic LLG equation we generated the random trajectories for
a sufficiently long time interval, until the magnetization components
reached the stationary regime. This procedure has been implemented many times in order
to construct an ensemble of the random solutions of the stochastic LLG equation \cite{sukhov}.
Each of these random solutions corresponds to a certain realization of the random noise,
while the statistical average over the ensemble of realizations designates the mean values
of the magnetization components.
Such average components of the magnetization vector ($\vec{m}$ and $\dot{\vec{m}}$) can be
used afterwards for the evaluation of spin current.
This numerical procedure is in general computationally expensive \cite{Vieira}, since
the number of realizations needed to reach a good accuracy of the solution to the stochastic
LLG equation is about one thousand. \cite{Etesami}.

\begin{center}
   \begin{figure}[!t]
    \centering
    \includegraphics[width=0.95\columnwidth]{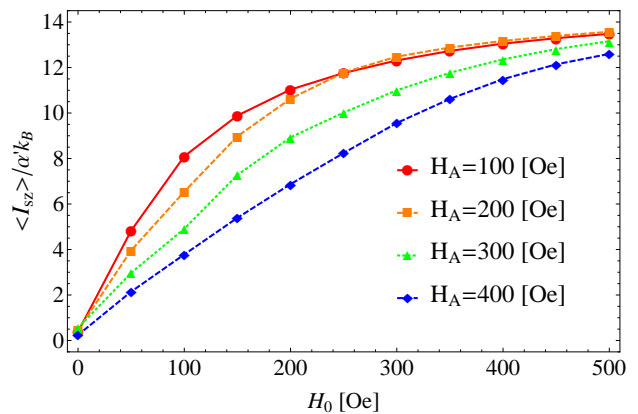}
        \caption{\label{fig6} The total spin current as a function of the external magnetic field $H_0$,
				obtained {\it via} numerical integration of the stochastic LLG equation for the indicated values of the positive
				anisotropy constant $H_A$ and for the parameters: $T_F^m=300$~K,
				$T_N=297$~K, $M_s=800$~G, $V=1.6\times10^{-18}$~cm$^3$, $\alpha=1.0$ and $\alpha'=0.05$.}
    \end{figure}
\end{center}
\begin{center}
   \begin{figure}[h]
    \centering
    \includegraphics[width=0.95\columnwidth]{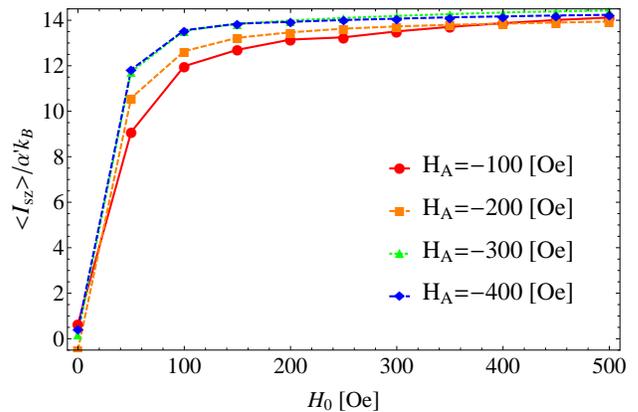}
        \caption{\label{fig7} Total spin current as a function of an external magnetic field $H_0$,
				obtained {\it via} a numerical integration of the stochastic LLG equation for the indicated values of the negative
				anisotropy constant, and for the parameters: $T_F^m=300$~K, $T_N=297$~K, $M_s=800$~G,
				$V=1.6\times10^{-18}$~cm$^3$, $\alpha=1.0$ and $\alpha'=0.05$.}
    \end{figure}
\end{center}
%
%

This numerical procedure can be used to calculate the spin pumping current from
the ferromagnetic part  to the  metallic one (first term in Eq.~(\ref{eq_1})).
In order to calculate the spin torque current
from the metal to the ferromagnet (second term in Eq.~(\ref{eq_1})), we have
to consider the random magnetic field in the metallic part, $\vec{h}'$, and
the corresponding correlation function, Eqs.~(\ref{eq.3}).
The metal temperature $T_N$ is interlaced with the strength of this stochastic field,
$2\alpha'k_BT_N/M_sV\gamma$.

Results of the numerical simulations are presented in Figs.~\ref{fig6} and \ref{fig7}.
As we see, the numerical results for the total spin current are very close to the results obtained
by means of the Fokker-Planck equation, see Fig.~\ref{fig1} and Fig.~\ref{fig2}. In both cases, the
total spin current increases with an external magnetic field.
In the case of a positive anisotropy field, the larger is the anisotropy, the
smaller is the spin current, while in the case of a negative anisotropy field, the spin current increases with increasing the
anisotropy field.

\section{Spin current in  $N|F|N$ structures}

\begin{center}
   \begin{figure}[!h]
    \centering
    \includegraphics[width=0.95\columnwidth]{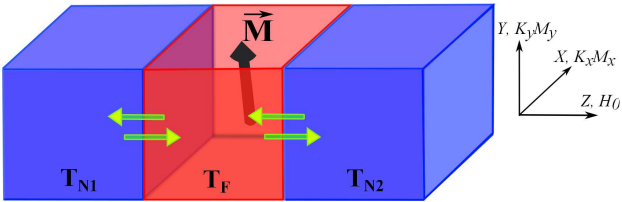}
        \caption{\label{skim2} Schematic presentation of the $N|F|N$ structure consisting of a magnetic element sandwiched between
        two nonmagnetic parts -- $N1$ on the left and $N2$ on the right side. Temperatures of the parts $N1$ and $N2$ are generally different. }
    \end{figure}
\end{center}

The technique used above can be also employed to calculate the thermally-induced spin current in a $N|F|N$ system,
shown schematically in Fig.~\ref{skim2}.
Now, the total spin current includes four terms,
\begin{equation}
\label{eq_10}
 \vec{I}_{s}=\vec{I}_{fl1}+\vec{I}_{sp1}+\vec{I}_{fl2}+\vec{I}_{sp2},
\end{equation}
where $\vec{I}_{sp1}$ and $\vec{I}_{sp2}$ stand for the spin pumping current from the ferromagnet
to the metallic parts $N1$ (left) and $N2$ (right), respectively (see Fig.~\ref{skim2}).
In turn, $\vec{I}_{fl1}$ and $\vec{I}_{fl2}$ stand for the spin torque current
flowing, respectively, from the left and right metallic parts to the ferromagnetic one due to thermal fluctuations.
We assume that the two metals  have generally different
temperatures, $T_{N1}$ and $T_{N2}$.

Upon laborious calculations, one finds the components of the spin pumping current flowing from the
ferromagnet  towards the two ($i=1,2$) metallic parts,
\begin{equation}
\label{eq_11}
\begin{split}
 & \langle I_{sp}^{(1)}\rangle=-\frac{M_{s}V}{\gamma}\alpha'\big(T_{N1}\big)f(T_{F}^{m}\big),\\
 & \langle I_{sp}^{(2)}\rangle=\frac{M_{s}V}{\gamma}\alpha'\big(T_{N2}\big)f(T_{F}^{m}\big),
 \end{split}
\end{equation}
where, for shortness, we introduced the following function:
\begin{equation}
\label{eq_13}
\begin{split}
&f(T_{F}^{m}\big)=\langle\omega_{z}^{\rm eff}\big(1-m_{z}^{2}\big)\rangle\\
&-\langle\omega_{x}^{\rm eff}m_{z}m_{z}\rangle-
\langle\omega_{y}^{\rm eff}m_{z}m_{y}\rangle .
 \end{split}
\end{equation}
In turn, for the spin torque components we find
\begin{equation}
\label{eq_12}
\begin{split}
 & \langle I_{fl}^{(1)}\rangle=2k_{B}\langle m_{z}\rangle\alpha'\big(T_{N1}\big)T_{N1},\\
 & \langle I_{fl}^{(2)}\rangle=-2k_{B}\langle m_{z}\rangle\alpha'\big(T_{N2}\big)T_{N2}.
 \end{split}
\end{equation}
As one can see from Eqs.~(13) and (14), the difference in the two components of the spin pumping current
transferred from the ferromagnetic part to the metals,
$\langle I_{sz}^{(1)}\rangle $ and $\langle I_{sz}^{(2)}\rangle$, is related to the temperature
dependence of the damping constants, or, more precisely, to the temperature dependence of the spin
conductance, $\alpha'\big(T_{N}\big)=\frac{\gamma\hbar}{4\pi M_{s}V}\, g_{r}\big(T_{N}\big)$.
Such a temperature dependence of the spin conductance $g_{r}\big(T_{N}\big)$ has been measured in
a recent experiment \cite{Joyeux}.

For convenience, we  denote
$\alpha'\big(T_{N1}\big)=\alpha'$ and $\alpha'\big(T_{N2}\big)=\alpha'+\Delta\alpha'$, and
rewrite the expression for the total spin current in the form
\begin{equation}
\label{eq_14}
\begin{split}
 & \langle I_{sz}\rangle = 2k_{B}\langle m_{z}\big(T_{F}^{m}\big)\rangle\alpha'\big(T_{N1}-T_{N2}\big)\\
 & +\frac{M_{s}V}{\gamma}f\big(T_{F}^{m}\big)\triangle\alpha'-2k_{B}\langle m_{z}\big(T_{F}^{m}\big)\rangle\Delta\alpha' T_{N2}.
 \end{split}
\end{equation}
If the difference between the temperatures of the  metals, $T_{N1}$ and $T_{N2}$, is not too
large, then the variation of the damping constant $\Delta\alpha'$
is very small, $\Delta\alpha'\ll \alpha'$.
In particular, the experimental data show the following change of the damping constant with temperature:
\cite{Joyeux} if $\Delta T= T_{N1}-T_{N2}\approx 350$K then $|\Delta\alpha'|/\alpha'\approx 0.28$.
For $T_{N1}-T_{N2}<100$ the relative variation of the damping constant is even smaller,
$|\Delta\alpha'|/\alpha'< 0.1$.
In such a case, the spin pumping currents transferred into the  metals
almost compensate each other, $\vec{I}_{sp1}\approx -\vec{I}_{sp2}$.
Thus, if the difference between the temperatures of the  metals, $T_{N1}$ and $T_{N2}$, is not high,
the $N|F|N$ system becomes nearly symmetric, and
only the spin torque current contributes to the total spin current.
The ferromagnetic part serves then as a conductor which transfers the spin current
from the hot metal to the cold one.
We note, that though the spin pumping currents to the left and right metals do not contribute to the total current,  $\vec{I}_{sp1}\approx -\vec{I}_{sp2}$, they  lead  to an enhanced Gilbert damping $\alpha'$ in  Eq.~(15).

Hence, in the symmetric case, the calculation of the total spin
current is greatly simplified, and the expression for the total spin current reads
\begin{equation}
\label{eq_15}
 \langle I_{sz}\rangle=2\alpha'k_{B}\langle m_{z}\rangle\big(T_{N1}-T_{N2}\big).
\end{equation}
As follows from Eq.~(\ref{eq_15}), the spin current flowing through the ferromagnet
mostly depends on the temperature difference between the metals. However, an additional temperature dependence enters through
the mean value of the
magnetization $\langle m_{z}\rangle$, which is a function of the magnon temperature
$T_{F}^{m}$.

In the absence of  magnetic anisotropy and in the limit of a weak external magnetic field,
$M_{s}VH_{0}<<k_{B}T_{F}^{m}$, the expression for the spin current takes the form
\begin{equation}
\label{eq_16}
\langle I_{sz}\rangle=\frac{2}{3}\alpha'M_{s}VH_{0}\frac{T_{N1}-T_{N2}}{T_{F}^{m}}.
\end{equation}
Obviously, the spin current decreases upon enhancing the magnon temperature.
From the physical point of view this result is rather clear. The magnetic part conducts the
spin torque current from the hot normal metal to the cold one, and the resistance of magnetic
element is increasing with the magnon temperature $T_{F}^{m}$.
Therefore, the spin current decreases at the elevated magnon temperature.

In the limit of a weak external magnetic field, $M_{s}VH_{0}<<k_{B}T_{F}^{m}$, the out-of-plane
magnetic anisotropy leads to the following correction in the spin current formula,
\begin{equation}
\label{eq_17}
  \langle I_{sz}\rangle=\frac{2}{3}\alpha'M_{s}VH_{0}
  \frac{T_{N1}-T_{N2}}{T_{F}^{m}}\bigg(1+\frac{2M_{s}VH_{A}}{15k_{B}T_{F}^{m}}\bigg).
\end{equation}
Thus, in the case of out-of-plane magnetic anisotropy, the spin current always decreases
with  increasing the magnon temperature $T_{F}^{m}$.
This result is clear since the magnetic anisotropy tends to align
the magnetic moment along the $z$ axis, and thus increases the spin
current in accordance with Eq.~(\ref{eq_15}), while the high magnon temperature $T_{F}^{m}$
reduces $\langle m_{z}\rangle$ and thus decreases the spin current.

In the case of in-plane magnetic anisotropy, the expression for the current reads
\begin{equation}
\label{eq_18}
 \langle I_{sz}\rangle=\frac{2}{3}\alpha'M_{s}VH_{0}\frac{T_{N1}-T_{N2}}{T_{F}^{m}}\bigg(1-\frac{2M_{s}VH_{A}}{15k_{B}T_{F}^{m}}\bigg).
\end{equation}
Thus, the temperature dependence is similar in both cases. The difference concerns the role of magnetic anisotropy -- the
out-of-plane anisotropy increases the spin current whereas the in-plane magnetic anisotropy reduces
the spin current.

\section{Numerical results for the $N|F|N$ structure}

As in the case of the $F|N$ structure, the analytical results derived above for the $N|F|N$ system can be supported by  direct numerical
integration of the stochastic LLG equation for the corresponding macrospin. Instead of this, we go in this section beyond the macrospin
approximation. It is well known, that the macrospin description breaks down for non-uniformly magnetized samples with the characteristic
lengths exceeding several tens of nm's. Beyond the macrospin formulation, the SSE effect
can be described by introducing the local magnetization
$\vec{m}\big(\vec{r},t\big)$. The description
can be then reduced to a discrete chain of magnetic moments.

As shown in recent work of Etesami et al.\cite{Etesami}, the spin current in the case of discrete
chain of magnetic moments can be calculated by using the following recurrent relations
\begin{equation}
\label{longitudinalspincurrent}
 \displaystyle I_{n}
=2Aa\sum_{l=1}^{n}m^{\mathrm{x}}_l(m^{\mathrm{y}}_{l-1}+m^{\mathrm{y}}_{l+1})
-m^{\mathrm{y}}_l(m^{\mathrm{x}}_{l-1}+m^{\mathrm{x}}_{l+1}),
\end{equation}
where $A$ is the exchange stiffness, $a$ is the unit cell size, $m_{i}^{x},~m_{i}^{x}$ are
the components of the individual magnetic moments, and $I_{n}$ is the site-dependent spin current.
We note that in the case of a non-uniformly magnetized sample, the spin current is not uniform
along the chain. The chain is oriented along the $x$ axes and the easy axes is in the $z$ direction.

The dynamics of magnetic moments $\vec{m}_{n}$ is described by coupled stochastic LLG
equations, we have to solve numerically. For more technical details we refer to the recent work
of Etesami et al.\cite{Etesami}.
The way how we take into consideration the interface effects is straightforward.
Namely, the contact of the magnetic chain with the left and right
metallic parts leads to modification of the damping constant for the first and last magnetic moments,
$\alpha_{1,N}=\alpha+\gamma \hbar g_{\mathrm{eff}}/(4\pi a M_s)$,
where $g_{\mathrm{eff}}$ is the effective spin-mixing constant at the interfaces. Constant $\alpha_{1,N}$ models enhancing of the Gilbert damping and is related to the spin pumping current. For more details see \cite{Kapelrud}.
The other point is that in the equations of motion for the first and last magnetic moments,
which are in contact with the left and right normal metals, include the spin torque term.

The idea of the spin torque is that it accounts for the effect of the interface on the adjacent magnetic
moments \cite{Etesami,Vieira}. We assume that the ratio between the amplitudes of the spin torque
terms is proportional to the ratio of the temperatures of the two metals,
$|\vec{I}^{\mathrm{in}}_\mathrm{L}|/|\vec{I}^{\mathrm{in}}_\mathrm{R}|=T^{N}_{L}/T^{N}_{R}$
and $\vec{I}^{\mathrm{in}}_\mathrm{L}=\lambda (T^{N}_{L},0,0)$ and $\vec{I}^{\mathrm{in}}_\mathrm{R}=\lambda (-T^{R}_{L},0,0)$, where $\lambda$ is a phenomenological constant.
\begin{center}
   \begin{figure}[!t]
    \centering
    \includegraphics[width=\columnwidth]{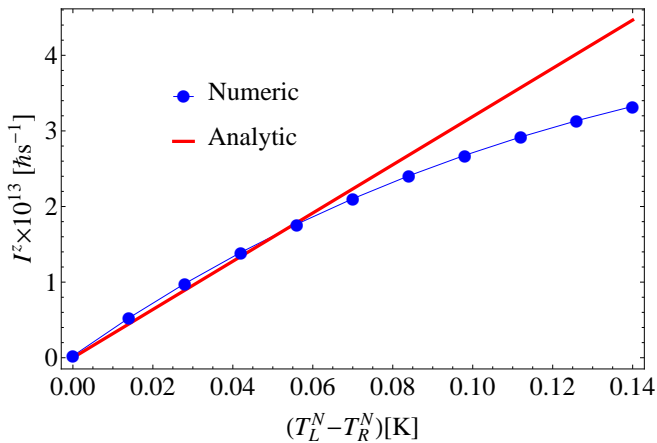}
        \caption{\label{fig8} Spin current through a FM cell as a function of the temperature difference of the two attached metals on the left(L) and right(R). The red curve is in line with analytical result  Eq.~(\ref{eq_17}), $\alpha'=0.05$ and $T^{m}=1 [K]$. The blue curve corresponds to the numerical result based on Eqs.~(\ref{longitudinalspincurrent}),(\ref{LLGN1}) and (\ref{LLGN2}). In the FM insulator spin torque is injected from the both left and right metals. Spin torque injected from the left metal is fixed $\lambda T_\mathrm{L}^N=5\times10^{15} [\hbar s^{-1}]$ while spin torque injected from the right metal is swaped. Here $\lambda=7.1\times10^{15} [\hbar s^{-1}T^{-1}]$ is a phenomenological constant which connects temperatures of the metals and corresponding spin torques $|\vec{I}^{\mathrm{in}}_\mathrm{L,R}|=\lambda T^{N}_{L,R}$. The spin-mixing coefficient at both interfaces (left and right) is $g_{eff}=1.14\times10^{22} [m^{-2}]$. The temperature of the FM system is $T_{F}=1 [K]$. The other parameters are presented in Table~\ref{tab2}.}
    \end{figure}
\end{center}

Thus, the equation of motion for the magnetic moment $\vec{m_{n}}$ can be written as
\begin{equation}
\label{LLGN1}
\begin{split}
  \displaystyle \frac{\partial \vec{m}_{n}}{\partial t} &=-\frac{\gamma}{1+\alpha_{n}^2}
	\left[\vec{m}_{n}\times \vec{H}^{\mathrm{eff}}_{n}\right]
  \\ & \quad-\frac{\gamma\alpha_{n}}{(1+\alpha_{n}^2)}
	\left[\vec{m}_{n}\times\left[\vec{m}_{n}\times \vec{H}^{\mathrm{eff}}_{n}\right]\right],
  \end{split}
\end{equation}
for $n=2,...N-1$, and
\begin{equation}
\label{LLGN2}
\begin{split}
  \displaystyle \frac{\partial \vec{m}_{1,N}}{\partial t} &=-\frac{\gamma}{1+\alpha_{1,N}^2}
	\left[\vec{m}_{1,N}\times \vec{H}^{\mathrm{eff}}_{1,N}\right]
  \\ & \quad-\frac{\gamma\alpha_{1,N}}{(1+\alpha_{1,N}^2)}\left[\vec{m}_{1,N}
	\times\left[\vec{m}_{1,N}\times \vec{H}^{\mathrm{eff}}_{1,N}\right]\right]
  \\ & \quad-\frac{\gamma}{M_{\mathrm{S}} a^3} \left[ \vec{m}_{1,N}
	\times \left[\vec{m}_{1,N} \times \vec{I}^{\mathrm{in}}_\mathrm{L,R} \right]\right]
  \\ & \quad-\frac{\gamma}{M_{\mathrm{S}} a^3}\beta \left[\vec{m}_{1,N}
	\times \vec{I}^{\mathrm{in}}_\mathrm{L,R}\right] ,
\end{split}
\end{equation}
for the magnetic moments in direct contact with the metallic parts.
The last two terms in Eq.~(\ref{LLGN2}) stand for the spin currents $\vec{I}^{\mathrm{in}}_\mathrm{L,R}$ injected from the metals
to the ferromagnetic chain. One of them is the torque of Slonczewski's type and the second one corresponds to an additional torque term \cite{Kapelrud, Etesami,Brataas}. The coefficient $\beta=0.001$ describes the relative strength of the last torque term with respect to the Slonczewski's torque.
\begin{table}[ht!]
\centering
\caption{The parameters used for the $N|F|N$ system in numerical simulations \label{tab2}}
\vskip0.5cm
\begin{tabular}{ll}
\toprule
DESCRIPTION & VALUE \\
\hline\\
Anisotropy constant $k$, [$\mathrm{J/m^3}$] \tabrule& $4.8\times 10^{4}$ \\
Exchange stiffness $A$, [J/m] \tabrule& $1.05\times 10^{-11}$ \\
Saturation magnetization $M_{\mathrm{S}}$, [A/m] \tabrule& $1.7\times 10^{6}$ \\
Gilbert damping $\alpha$ \tabrule& $0.01$ \\
FM cell size $a$, [m] \tabrule& $20\times 10^{-9}$ \\
\textbf{External field $H_0$, [T]} \tabrule& \textbf{$10^{-5}$} \\
\textbf{Number of realizations, $R$} \tabrule& \textbf{$100$} \\
\bottomrule
\end{tabular}
\end{table}

The results of numerical calculations are shown in the Fig.~\ref{fig8}. We plotted the spin current
conducted through the FM insulator. In numerical calculations, the FM insulator is modeled by a chain of coupled magnetic
moments. The spin current is plotted as a function of temperature difference between two metals $T^{N}_{L}-T^{N}_{R}$. As we see numerical result fits the analytical behavior (Eq.~(\ref{eq_17})). Until the difference between metal temperatures is not too large, the spin current increases linearly with the difference $T^{N}_{L}-T^{N}_{R}$.

\section{Conclusions}

We have studied the influence of in-plane magnetic anisotropy on the
thermally activated spin current flowing through interfaces between ferromagnetic insulator and
nonmagnetic metal. We have considered two different systems: (i) ferromagnetic insulator and nonmagnetic metal in a direct contact, $N|F$, and (ii) ferromagnetic  insulator sandwiched between two nonmagnetic metals, $N|F|N$ structure.
In the symmetric case, $K_x=K_y$, we  derived  analytical expressions for the average
spin current in several limiting situations.

In the case of a weak anisotropy and a weak external field, $M_sVH_0/k_BT_F^m<1$
and $H_{A}M_sV/2k_BT_F^m<1$ (this case refers to the high magnon temperature $T_F^m$),
the total spin current is given by the formula $\langle I_{sz}\rangle=2\alpha'
k_B\frac{M_sVH_0}{3k_BT_F^m}\left(T_F^m-T_N\right)\left(1-\frac{2}{3}\frac{H_{A}M_sV}{2k_BT_F^m}\right)$.
The in-plane positive anisotropy, $H_{A}>0$, suppresses then the spin
current, while the negative anisotropy, $H_{A}<0$, enhances the spin current.

In the case of a strong magnetic field and a weak anisotropy,
$\frac{1}{2}H_{A}M_sV<<k_BT_F^m<<H_0M_sV$, we identified  a different regime,
and the expression for the spin current reads then
$\langle I_{sz}\rangle=2\alpha' k_B\big(T_F^m-T_{N}\big)
\bigg(1-\frac{3H_{A}}{H_0}\frac{k_BT_F^m}{M_sVH_0}\bigg).$
In this case, the magnetic anisotropy suppresses the spin current, too.
The situation is different in the asymmetric
case of $K_x\neq K_y$. We find that for a weak magnetic field, $2H_0/\left(K_x+K_y\right)<1$,
the asymmetry reduces the spin  current, see Fig.~\ref{fig5}.

Another conclusion is  that if the difference between
temperatures $T_{N1}$ and $T_{N2}$ of the two metals  in the $N|F|N$ system is not too large,
the problem becomes effectively symmetric.
The spin pumping currents flowing in  opposite directions
compensate then each other, $\vec{I}_{sp1}=-\vec{I}_{sp2}$, so that only the
spin-torque components contribute to the total spin current.
In this case, the ferromagnetic element of the structure serves as a conductor,
which conducts the spin-torque current from the hot to cold metal.
In the linear response regime,  the spin current is linear in
the temperature difference $(T_{N1}-T_{N2})$. If the difference between the temperatures of two
metals is large enough, this breaks the symmetry between the spin pumping currents,
$\vec{I}_{sp1}\neq-\vec{I}_{sp2}$ driving the system beyond the linear response regime.

\section*{Acknowledgements}

This work has been partly supported by the National Science Center in Poland by the Grant
DEC-2012/04/A/ST3/00372 (VKD and JB) and the DFG through SFB762.

\begin{appendix}

\section{Average values of the magnetization components}

The mean components of the magnetization, see Eq.~(\ref{eq_6}), are given as
\begin{equation}
\label{eq_A2}
\langle m_z\rangle=\frac{2\pi}{Z}e^{B}\,\int\limits_{-1}^{+1}dx\, x
 e^{Ax-Bx^2}I_0(\zeta),
\end{equation}
\begin{equation}
\langle1-m_z^2\rangle=\frac{2\pi}{Z}e^{B}\int\limits_{-1}^{+1}dx \left(1-x^2\right)
e^{Ax-Bx^2}I_0(\zeta),
\end{equation}
\begin{equation}
\langle m_zm_x^2\rangle=\frac{\pi}{Z}e^{B}\int\limits_{-1}^{+1}dx\, x \left(1-x^2\right)
e^{Ax-Bx^2}[I_0(\zeta)+I_1(\zeta)],
\end{equation}
\begin{equation}
\langle m_zm_y^2\rangle=\frac{\pi}{Z}e^{B}\int\limits_{-1}^{+1}dx\, x \left(1-x^2\right)
e^{Ax-Bx^2}[I_0(\zeta)-I_1(\zeta)],
\end{equation}
where
\begin{equation}
\label{eq_A6}
Z= 2\pi e^{B}\int\limits_{-1}^{+1}dx\,e^{Ax-Bx^2}I_0(\zeta)
\end{equation}
and $\zeta =\Delta\left(1-x^2\right)$, while  $I_0(\cdot\cdot\cdot)$ and
$I_1(\cdot\cdot\cdot)$ are the modified Bessel functions of the
zeroth and first order, respectively.
Apart from this, we introduced here the following notation: $B=\frac{1}{2}(B_x+B_y)$ and
$\Delta=\frac{1}{2}(B_x-B_y)$.

In case of a weak magnetic anisotropy, $B_x=B_y=B<1$, the spin current and the mean components of the magnetization  take the form
\begin{equation}
\label{eq_A7}
 \begin{split}
&\langle I_{sz}\rangle=2\alpha'k_B\big(T_F^m-T_N\big)  \\
&\times\Big\{L(A)-2B\Big(\frac{3L(A)}{A^2}+\frac{L^2(A)}{A}-\frac{1}{A}\Big)\Big\}
 \end{split}
\end{equation}

\begin{equation}
\langle m_z\rangle=L(A)-2B\Big(\frac{3L(A)}{A^2}+\frac{L^2(A)}{A}-\frac{1}{A}\Big),
\end{equation}

\begin{equation}
\langle1-m_z^2\rangle=\frac{2L(A)}{A}-4B\Big(\frac{6L(A)}{A^3}+\frac{L^2(A)}{A^2}-\frac{2}{A^2}\Big),
\end{equation}

\begin{equation}
\begin{split}
\label{eq_A10}
&\langle m_z(m_x^2+m_y^2)\rangle=-\frac{6L(A)}{A^2}+\frac{2}{A}\\
&+4B\bigg(\frac{30L(A)}{A^4}+\frac{L(A)}{A^2}+\frac{3L^2(A)}{A^3}-\frac{10}{A^3}\bigg) ,
\end{split}
\end{equation}
where  $L(A)=\coth(A)-\frac{1}{A}$ is the Langevin function.

\end{appendix}

\end{document}